\documentstyle[12pt]{article}
\begin{document}{}
\newtheorem{guess}{Conjecture}
\newtheorem{proovit}{Theorem}
\newtheorem{helpme}{Lemma}
\newtheorem{shoutit}{Statement}
\newtheorem{defit}{Definition}
\begin{titlepage}
\begin{flushright}
UBCTP--92--016 preprint \today\\[0.25in]
\end{flushright}
\begin{center}
{\Huge Generalized embedding variables\\
for geometrodynamics and\\
spacetime diffeomorphisms:\\
Ultralocal coordinate conditions}
{\\[0.5in] \Large  Stephen P\@. Braham}
{\\[0.5in] \em Department of Physics
\\University of British Columbia
\\Vancouver, BC, Canada V6T 1Z1
\\E-mail: braham@physics.ubc.ca\\[0.5in]}
\end{center}
\centerline{\large\bf Abstract}
\vskip 0.25in
{\small
We investigate the embedding variable approach to
geometrodynamics advocated in work
by Isham, Kucha\v{r} and Unruh for a general class of coordinate
conditions that mirror the Isham-Kucha\v{r} Gaussian condition but
allow
for arbitrary algebraic complexity. We find that the
same essential structure present in the ultralocal Gaussian condition
is
repeated in the general case. The resultant embedding--extended
phase space contains a full
representation of the Lie algebra of the spacetime diffeomorphism
group as well as a consistent pure gravity sector.\\[0.15in]
PACS $04.60.+{\rm n}$, $04.20.{\rm Fy}$, $02.40.+{\rm m}$,
$11.10.{\rm Ef}$
}
\end{titlepage}

\section{Introduction}{}\label{intro}
The central beauty of general relativity, general covariance, is both
a joy and a curse for those who have chosen to work in the field of
geometrodynamics. To study dynamics, we need some form of
preferred, God given, time coordinate,
but we know that a theory that is invariant under spacetime
diffeomorphisms
cannot have an externally preferred time.
In fact, as we shall demonstrate in the
next section, fixation of the coordinates  in the
Einstein-Hilbert action prior to the variation of the
dynamical variables (i.e. certain
combinations of the components of the metric tensor) does not lead to
the full Einstein equations, and thus, at a first glance, it seems
that a specification of time will be impossible
if we hope to recover general relativity.
Furthermore, gravity possesses a far richer
concept of time than that which our pre-relativistic ancestors grew up
with; the natural structure for time in Lorentzian signature
geometrodynamics is the set of {\it all} spacelike hypersurfaces in a
given spacetime, dubbed {\it hypertime} by Kucha\v{r} \cite{KK4}.
Hypertime is
many-fingered; every point in the spacetime can experience time at its
own rate, corresponding to the various ways of
pushing forward a hypersurface into the future,
and thus time is not some single variable, as in conventional
dynamics, but instead can only be represented by the full spacetime
coordinates $X^{\mu} (t, x)$ of a chosen hypersurface in a foliation
of the spacetime, where $X^{\mu}$ denotes a general coordinate on the
spacetime manifold, and $x$ denotes a position on the hypersurface,
which is selected from the foliation by the label $t$. The nature of
hypertime becomes a major problem when we try to quantize gravity via
canonical methods, as discussed by Kucha\v{r} \cite{KK4}
as well as Unruh and Wald \cite{BILLBOB1}. This is obvious
from the structure of the Wheeler-DeWitt equation (cf.~Section
\ref{hamX})
which has no manifest time parameter and thus no easy interpretation in
terms of quantum dynamics, leading to the so-called
problem of time. In ground breaking work, Isham, Kucha\v{r}
and Unruh \cite{DIFIK1,DIFIK2,BILLUM1} have taken a bold step
into land that most relativists fear to
tread, and have indeed advocated the introduction of a preferred time
coordinate into geometrodynamics. Their work has indicated the process
by which we may do this and still recover normal Einsteinian gravity.
The Isham-Kucha\v{r}-Unruh (IKU) approach involves
adding a set of variables to
gravity that represent a preferred coordinate system, bolted firmly in
place to
the gravitational field by a set of coordinate conditions. These
variables
can be considered to be specifications of the embedding of
hypersurfaces
into the spacetime, in the sense of the $X^{\mu}$ described earlier,
and thus are a representation of the hypertime concept. The embedding
variables are appended to the Einstein-Hilbert action by using a
Lagrange
multiplier term involving the desired coordinate
conditions (cf. next section). The multipliers can
then be removed as free
variables at the Hamiltonian level by some construction, giving an
extended
phase space consisting of only the gravitational degrees of freedom
plus the embedding variables.
The embedding-extended theory can be regarded as one describing a
matter
field coupled to Einsteinian geometrodynamics. For
the Gaussian \cite{DIFIK2,KT1} and
harmonic \cite{KT2,SK1} coordinate conditions, it can be
demonstrated that one can
consistently impose extra constraints on the theory to obtain standard
general relativity {\it in vacua}. Furthermore, the extended phase
space contains a full representation of the Lie algebra of the
spacetime
diffeomorphism group. However, all work so far has used specific
coordinate conditions, and thus a general understanding of the
dependence
of the resulting formalism on the conditions themselves has
been lacking.

With the former in mind, it seems apt to embark on a program in
which we try to formulate extremely general coordinate conditions, so
that
we may understand the general nature of the embedding approach. In this
paper, we will study embedding variables corresponding to coordinate
conditions that can be written in terms of ultralocal algebraic
functions
of the metric with respect to the embedding on a spatial hypersurface.
This condition includes the Isham-Kucha\v{r} Gaussian condition, the
Unruh unimodular condition \cite{BILLUM1}
(with added frame choice conditions) as well
as the conformal gauge condition \cite{KT3} (in two dimensions),
but does not include the harmonic condition.
We shall verify the properties discovered for the Gaussian coordinate
condition and shall find that the general algebraic
complexity of the description does not damage
the geometrical content of the resulting Hamiltonian formulation. We
will briefly discuss issues related to the use of this method for
canonical quantization of the gravitational field, but will primarily
focus on the purely classical geometrodynamical content of the
construction.
\section{Coordinate conditions in the Lagrangian}{}\label{lag}
We will start by briefly examining the effect of using coordinate
conditions in gravity in the manifestly covariant Lagrangian
formalism before proceeding, in the next section, with the
Hamiltonian analysis, which shall form the major part of this
paper. We will assume for the entire paper that the spacetime
$n$-manifold, $\cal M$, is of the form $M \times {\cal T}$, where
$M$ is an $(n-1)$-manifold without boundary,
and $\cal T$ is a 1-manifold (possibly with boundary). We shall further
assume that $\cal M$ possesses a metric ${\bf g}$ of Lorentzian
signature, giving it the structure of a pseudo-Riemannian manifold.
Such
a metric is a map defined in terms of the tangent space at a point
$Y \in {\cal M}$,
\begin{equation}{}\label{gdef}
{\bf g}_Y: T_Y ({\cal M}) \times T_Y ({\cal M}) \rightarrow {\Re}.
\end{equation}
The foliation embedding of $M$ into $\cal M$, $i(t): M \rightarrow
{\cal M}$,
will be restricted to be
everywhere spacelike for all values of $t \in {\cal T}$. We can then
define the (positive definite) metric tensor on each leaf of the
foliation by
\begin{equation}{}\label{hdef}
{\bf h}_{(t, x)} =  i^* {\bf g}_{Y(t, x)},
\end{equation}
where $x \in M$, ${\bf h}: T_x (M) \times T_x (M) \rightarrow {\Re}$,
and
$Y(t,x) = i(t)x$.

We can work in a coordinate patch of $\cal M$, $U \times V$,
where $U \subseteq M$ and $V \subseteq {\cal T}$. We will denote
the resulting coordinates by $x^a(Y)$ and $t(Y)$, where lower
case Latin characters range
from one to $n-1$ and will denote the n-dimensional coordinate
$(t, x^a)$ by $x^{\mu}$, with lower case greek letters running from
zero to $n-1$. We can use the push-forward, $Y_*$, derived from
the embedding to isomorphically map $T_{(t,x)}(M \times {\cal T})$
to $T_Y ({\cal M})$, and shall therefore not worry about distinguishing
between objects on the various bundles constructed over the manifold in
either language. The condition that the embedding be
spacelike and that ${\bf g}$ be Lorentzian is that
for all points $Y \in {\cal M}$ we have
\begin{equation}{}\label{glor1}
{\bf g}_Y \left( v^a \frac{\partial}{\partial x^a},
	v^b \frac{\partial}{\partial x^b}
       \right) > 0,
\end{equation}
for all $v^a \in {\Re}^{n-1}$ with $v^a \not = 0$, and
\begin{equation}{}\label{glor2}
\ {\bf g}_Y \left( t^\mu \frac{\partial}{\partial x^\mu},
\frac{\partial}{\partial x^a}
       \right) =0,\ \forall a \Rightarrow
\ {\bf g}_Y \left( t^\mu \frac{\partial}{\partial x^\mu},
t^\mu \frac{\partial}{\partial x^\mu}
       \right) < 0,
\end{equation}
where $t^\mu \in {\Re}^n$, with $t^\mu \not = 0$.
We can express the metric in terms of these coordinates by
\begin{equation}{}\label{metco}
{\bf g}_Y = g_{\mu \nu}(Y)\ dx^{\mu} \otimes dx^{\nu},
\end{equation}
and similarly for ${\bf h}$ in its coordinate system, as well as other
objects.
The standard Einstein-Hilbert action for general
relativity is then
\begin{equation}{}\label{EHdef}
S^G \left[ {\bf g} \right] =
	    \int_{\cal M} dt \, d^3 x \sqrt{-g} R\left[{\bf g}, Y
	    \right)
	    - 2 \int_{\partial {\cal M}}  d^3 x \sqrt{h} K\left[{\bf
	    g},
	    x \right),
\end{equation}
where $R\left[ {\bf g}, Y \right)$ is the curvature scalar of
${\bf g}$, $K\left[ {\bf g}, x \right)$ is the
trace of the extrinsic curvature tensor of ${\bf g}$ relative to the
embedding of the boundary, $\partial {\cal M} = M \times \partial {\cal
T}$,
and $g$ and $h$ are the determinants of the metrics in the $x^{\mu}$
and $x^a$ coordinate systems respectively. The style of brackets is a
reminder
of the {\it functional} dependence on the metric.
We then write this in terms of a gravitational
Lagrangian, $L^G \left[ {\bf g} \right]$, by defining
\begin{equation}{}\label{Glag}
S^G \left[ {\bf g} \right] = \int_{\cal T} dt L^G \left[ {\bf g}, t
\right),
\end{equation}
where $L^G$ only depends on $K$, $R$ and ${\bf g}$ on the leaf
of the foliation labeled by $t$. We can now define the usual Einstein
tensor by
\begin{equation}\label{Eind}
G_{\mu \nu} \left[ {\bf g}, Y \right) = \frac{1}{\sqrt{-g(Y)}}
			   \frac{\delta}{\delta g^{\mu \nu}(Y)}
			   S \left[ {\bf g} \right].
\end{equation}
where $g^{\mu \nu}$ is the standard inverse to ${\bf g}$, expressed
in terms of the $x^{\mu}$ coordinate system. The resulting Einstein
equations are, as always
\begin{equation}{}\label{Eineq}
G_{\mu \nu} = 0,
\end{equation}
as long as we can freely vary $g^{\mu \nu}$, which corresponds to
varying
the embedding, $i$, as well as the metric, ${\bf h}$, on each leaf.

However, what happens if we wish to keep the embedding fixed, and work
in
a given coordinate system? In this case, $g^{\mu \nu}$ cannot be
varied freely, and we will not obtain the full Einstein equations.
Consider
a general coordinate system $(U_\alpha, X_{(\alpha)})$, where
${\cal M} = \bigcup U_\alpha$ and $X_{(\alpha)}$ takes $U_\alpha$
injectively into ${\Re}^n$, with image $X_{(\alpha)}^{\mu}$ and
with the map $X_{(\alpha)} X_{(\beta)}^{-1}$
being a $C^\infty$ diffeomorphism on $X_{(\beta)} (U_\alpha
\cap U_\beta)$; For this paper, we will define a {\it preferred}
coordinate system to be one in which we choose `almost one'
$X_{(\alpha)}$
in the following sense: the only maps allowed in the preferred atlas
are a subset with $dX_{(\alpha)} = dX_{(\beta)}$ in $U_\alpha \cap
U_\beta$.
Therefore, the transition function derivatives,
\begin{equation}{}\label{tfd}
X_{\nu (\alpha)}^{\mu} = \frac{\partial X_{(\alpha)}^{\mu}}{\partial
x^\nu},
\end{equation}
between an arbitrary coordinate system $x^\nu$ and $X_{(\alpha)}^\mu$,
are independent of $\alpha$. We will denote this common value by
$X^{\mu}_{\nu}$ and denote the images of the coordinate maps by
$X^{\mu}$,
and call these variables {\it embedding variables},
from now onwards. It is important to note that such a
coordinate system will usually be singular.
To select the $n$ $X^{\mu}$'s, we need $n$ independent
conditions on the components of ${\bf g}$ in the $X^{\mu}$
system, $\bar{g}^{\mu \nu}$,
\begin{equation}{}\label{conds}
F^A \left( \bar{g}^{\mu \nu}, Y \right) = 0,
\end{equation}
where
\begin{equation}{}\label{gbd}
\bar{g}^{\mu \nu}(Y) = {\left[ {\bf g}_Y \left(
		       \frac{\partial}{\partial X^{\mu}},
		       \frac{\partial}{\partial X^{\nu}} \right)
		       \right] }^{-1} =
		       X^{\mu}_{\rho}(Y)
		       X^{\nu}_{\sigma}(Y) g^{\rho \sigma} (Y).
\end{equation}
Here, $A$ runs from zero to $n-1$ and $g^{\mu \nu}$ is, as before,
the inverse metric in the $x^\mu$ coordinate system.
We will restrict ourselves to {\it ultralocal} coordinate
conditions, which are defined as follows:
\begin{defit}[Ultralocal coordinate conditions]{}\label{ucdef}
An ultralocal coordinate condition is a specification of
$n$ independent conditions, $F^A \left( \bar{g}^{\mu \nu} \right) = 0$,
with the property that $F^A \left( \bar{g}^{\mu \nu} \right)$,
for each value of $A$, is an ultralocal function of only the induced
metric, $\bar{g}^{\mu \nu}$.
We will further require that each $F^A \left( \bar{g}^{\mu \nu}
\right)$
is a smooth function of $\bar{g}^{\mu \nu}$ everywhere, except possibly
for a
discrete set of metrics.
\end{defit}
Notice that we have two important invariant
behaviors here; $\bar{g}^{\mu \nu}$ is invariant to changes of
$x^{\mu}$
coordinates as well as to the choice of $\alpha$ in
$X^{\mu}_{(\alpha)}$.
Thus, each $F^A$ transforms as a scalar under changes of the $x^\mu$
coordinates.

Now, we can use the standard trick to calculate the variation
of $S^G$ whilst maintaining the conditions given to us by
Equation~(\ref{conds}); we append a Lagrange multiplier term to $L^G$,
\begin{equation}{}\label{newL}
L(t) = L^G(t) + L^R(t),
\end{equation}
with
\begin{equation}{}\label{refL}
L^R(t) = -\frac{1}{2} \int_{M} d^3x \sqrt{-g(t,x)} \lambda_A (t,x) F^A
	  \left( \bar{g}^{\mu \nu}, Y(t,x) \right).
\end{equation}
We then make the following obvious definitions:
\begin{equation}{}\label{newS}
S \left[ {\bf g} \right] = S^G \left[ {\bf g}
    \right] +S^R \left[ {\bf g} \right]
    ,\ S^R \left[ {\bf g} \right] = \int_{\cal T} dt
    L^R \left[ {\bf g}, t \right).
\end{equation}
Variation
of $\lambda_A$ then gives us Equation~(\ref{conds}), and we may fix
$\lambda_A$ by varying $X^\mu$. Notice that we have used
general covariance to write the integration in the Lagrange multiplier
term as one over the $x^a$ variables. Similarly, we can also write
$L^G$ purely in terms of $x^{\mu}$ objects, and thus the $X^{\mu}$'s
only appear through the $\bar{g}^{\mu \nu}$ term in the conditions. It
is then a simple matter to see that our
new equations of motion corresponding to
variation of $g^{\mu \nu}$ are
\begin{equation}{}\label{newE}
G_{\mu \nu} = T_{\mu \nu},
\end{equation}
where
\begin{equation}{}\label{EMT}
T_{\mu \nu}(t,x) = - \frac{1}{\sqrt{-g(t,x)}} \frac{\delta}
		{\delta g^{\mu \nu}(t,x)} L^R(t).
\end{equation}
Now, $T_{\mu \nu}$ will transform as a rank two covariant tensor
under general transformations of the $x^\mu$ coordinates.
Furthermore, it is well-known that diffeomorphism invariance is
revealed through the contracted Bianchi identities $G_{\mu \nu}^{;\nu}
= 0$,
which force the solutions to obey $T_{\mu \nu}^{;\nu} =0$. These
relations give us differential equations for
$\lambda_A$ which are first order in time. However, we see that these
are exactly the equations we get from the variation of $X^\mu$ from
the following theorem:
\begin{proovit}[$X^\mu$ and the Bianchi identities] The variation of
the embedding variables, $X^\mu$, in the extended
action generates the contracted
Bianchi identities for the energy-momentum tensor, $T_{\mu \nu}$.
\end{proovit}
{\bf Proof:} Now, $T_{\mu \nu}^{;\nu}$ is a covector
and thus we can
evaluate it in any coordinate system and
foliation; Let us choose $x^\mu = X^\mu$. We
then have $\bar{g}^{\mu \nu} = g^{\mu \nu}$ and
\begin{equation}{}\label{vary}
\delta \bar{g}^{\mu \nu} = 2 {\delta X}^{(\mu;\nu)}.
\end{equation}
Thus we have
\begin{equation}{}\label{varyS}
\delta S = \delta S^R = 2 \int_{\cal M} dt \, d^3x \left( \frac{\delta}
			{\delta g^{\mu \nu}(Y)} S^R \right)
			{\delta X}^{(\mu;\nu)}(Y).
\end{equation}
Therefore, integration by parts gives us
\begin{equation}{}\label{varyS2}
\delta S = 2 \int_{\cal M} dt \, d^3x \sqrt{-g(Y)} T_{\mu
\nu}^{;\mu}(Y)
	      {\delta X}^{\nu}.
\end{equation}
Hence we arrive at our final destination,
\begin{equation}{}\label{finalstep}
\frac{\delta S}{\delta X^{\nu}(Y)} = 0 \Leftrightarrow T_{\mu
\nu}^{;\mu}
	      (Y) =0,
\end{equation}
which is then true in all coordinate systems by general covariance.
{\bf QED}

Thus the action of spacetime diffeomorphism is now contained within the
extended Lagrangian, $L$. This procedure allows us to consider the
$X^{\mu}$
variables as {\it scalar matter fields} on $\cal M$ with
a conserved energy-momentum tensor, with the caveat that they are
to have the multivalued property discussed above. These fields may
or may not behave like physical matter fields, as has been discussed
elsewhere \cite{KT1,KT2}.

To finish this section, we will briefly examine
some possible ultralocal coordinate
conditions:
\subsection{Gaussian condition}
Here, for any dimension $n$, we pick the Gaussian time condition,
\begin{equation}{}\label{GTC}
F^0({\bar{g}}^{\mu \nu}) = {\bar{g}}^{00} +1,
\end{equation}
and the Gaussian frame condition,
\begin{equation}{}\label{GFC}
F^i ({\bar{g}}^{\mu \nu}) = {\bar{g}}^{0i}.
\end{equation}
This condition forms the basis for the original work by Isham and
Kucha\v{r} \cite{DIFIK2}. The solutions are metrics of the block form
\begin{equation}{}\label{GCform}
\bar{g}^{\mu \nu}(Y) = \left( \begin{array}{cc}
		       -1 & 0 \\
		       0 & \gamma^{ij}(Y) \end{array} \right),
\end{equation}
where $\gamma^{ij}$ is the metric on the $X^0=$ {\it constant}
surfaces, with respect to the $X^a$ coordinate system.

\subsection{Conformal gauge condition}
This is a condition that we see primarily in string theory, in
$n=2$. We have
\begin{equation}{}\label{CGC}
F^0(\bar{g}^{\mu \nu}) = \bar{g}^{12},\ F^1(\bar{g}^{\mu \nu}) =
			\bar{g}^{11} + \bar{g}^{22}.
\end{equation}
This puts the metric into conformal form,
\begin{equation}{}\label{CCform}
\bar{g}^{\mu \nu} (Y) = \Omega^2 (Y) \left( \begin{array}{cc}
			-1 & 0 \\
			 0 & 1 \end{array} \right),
\end{equation}
where $\Omega$ is the conformal factor.
This condition has been studied by Kucha\v{r} and Torre \cite{KT1}.

\subsection{Conditions of unimodular type}
Here we will introduce a generalization of the unimodular condition
proposed by Unruh \cite{BILLUM1}, which corresponds to,
\begin{equation}{}\label{UTC}
F^0(\bar{g}^{\mu \nu}) = \det ( \bar{g}^{\mu \nu} ) +1.
\end{equation}
We will call this condition the {\it unimodular time condition}.
However, Equation~(\ref{UTC}) does not provide enough information
to specify all $n$ embedding variables, as has
been discussed by Kucha\v{r} \cite{KK4}. To fully encode hypertime,
we will need extra {\it frame} conditions.
For each choice of frame condition, we will have a corresponding
generalization of the unimodular condition.
One simple choice is to
use the Gaussian frame conditions given by Equation~(\ref{GFC}),
giving us a metric of the form
\begin{equation}{}\label{UTform}
\bar{g}^{\mu \nu}(Y) = \left( \begin{array}{cc}
		       -{\det}^{-1} (\gamma^{ij}(Y)) & 0 \\
		       0 & \gamma^{ij}(Y) \end{array} \right),
\end{equation}
where $\gamma^{ij}$ has the same interpretation as before. Again, this
general class of conditions will work for any value of $n$.
\section{Embeddings in the Hamiltonian formalism}{}\label{hamX}
We will now proceed to our main purpose; for geometrodynamics,
we are mainly interested in the Hamiltonian evolution of the theory.
Let
us start by looking at pure Einstein gravity, without extra variables.
We
will use the standard ADM technique \cite{ADM}, in which
we write the inverse of
the action of $i^*$ on ${\bf g}$ as
\begin{equation}{}\label{admmet}
{\bf g}_Y =  -(N^2-N_i N^i) dt \otimes dt + N_i dt \otimes d x^i
	    + N_i dx^i \otimes dt +h_{ij} dx^i \otimes dx^j,
\end{equation}
in terms of the $x^{\mu}$ coordinates. $N$ is the standard lapse
function and $N^i \in T(M)$ is the standard shift vector field. If we
now use an overdot to denote differentiation with respect to the
$t$ coordinate, then we can define the momenta conjugate to the
$h_{ij}$ by
\begin{equation}{}\label{momh}
{\pi}^{ij}(t,x) = \frac{\delta L^G(t)}{\delta {\dot h}_{ij} (t,x)}.
\end{equation}
The lapse and shift have no conjugate momenta, and thus are not
true dynamical variables. Thus we can
define coordinates $(h_{ij}, {\pi}^{ij})$ on the
actual gravitational phase space $T^*({\cal C}_h)$, the cotangent
bundle over
the space of positive definite metrics on $M$, ${\cal C}_h$.
It is in this space that the usual dynamics of gravity take place.
However, the action in these coordinates is now
\begin{equation}{}\label{hamac}
S^G [h_{ij}, \pi^{ij}, N, N^i] = \int_{\cal M} dt \, d^3x \left(
	     {\pi}^{ij} {\dot h}_{ij} - N H^G - N^i H^G_i \right),
\end{equation}
and where the superhamiltonian is given by
\begin{equation}{}\label{superh}
H^G = \frac{1}{\sqrt{h}} \left( {\pi}_{ij} {\pi}^{ij}
			- \frac{1}{2} {\pi}^i_i {\pi}^j_j \right)
			- \sqrt{h} R[{\bf h}],
\end{equation}
where the indices are raised and lowered by using $h_{ij}$,
and the supermomentum is given by
\begin{equation}{}\label{superm}
H^G_i = -2 {\pi}_{i;j}^j.
\end{equation}
Now,  variation of the redundant lapse
and shift functions produces constraints on the
dynamical system, given by
\begin{equation}{}\label{constr}
H^G = 0 =H^G_i.
\end{equation}
The constraints reduce the arena of geometrodynamics down to the
constraint manifold $\Gamma \subset T^*({\cal C}_h)$ on which
Equations~(\ref{constr}) hold. Ideally one would like to then solve
these constraints and thus produce a standard, unconstrained, phase
space on which normal dynamical evolution would take place. However,
the
constraints are by no means easy to solve and there may even be
topological obstructions to a general solution \cite{TOP1PETR}.
Another problem with
these constrains is their canonical commutation relations, which are
\begin{equation}{}\label{ccr1}
\{ H^G(x), H^G(x')\} = \left( h^{kl} H^G_k (x) - h^{kl}H^G_k (x')
\right)
		      {\delta}_{,l} (x,x'),
\end{equation}
\begin{equation}{}\label{ccr2}
\{ H^G_k(x), H^G(x')\} = H^G(x) {\delta}_{,l} (x,x'),
\end{equation}
and
\begin{equation}{}\label{ccr3}
\{ H^G_k(x), H^G_l(x')\} = H^G_l (x) {\delta}_{,k} - H^G_k (x')
{\delta}_{,l}.
\end{equation}
Our problem with these relations is the occurrence of $h^{kl}$ in
Equation~(\ref{ccr1}); This makes it impossible to construct a
representation
in $T^*({\cal C}_h)$ of the Lie algebra, ${\rm LDiff}({\cal M})$, of
the
the Lie group of spacetime diffeomorphisms, ${\rm Diff}({\cal M})$,
which is the gauge group of general relativity (cf. Ref.
\cite{DIFIK2} for a discussion).
Our third problem arises if we do not wish to solve the constraints and
proceed to quantization; In this case, the constraints becomes
operators,
${\hat{H}}$ and ${\hat{H}}_i$, on the wavefunction, $\psi ({\bf h})$.
The equation ${\hat{H}}_i \psi = 0$ then tells us that $\psi$ is
invariant
under the action of ${\rm Diff}(M)$, the group of spatial
diffeomorphisms,
and is thus only a function of 3-geometries, ${}^{(3)} {\cal G}$,
with the superhamiltonian constraint becoming the Wheeler-DeWitt
equation,
\begin{equation}{}\label{wdw}
{\hat{H}} \psi ( {}^{(3)} {\cal G} ) = 0.
\end{equation}
This equation has no manifest time parameter, thus producing extreme
difficulties in the interpretation of quantum geometrodynamics (QGD).
Reduction (solving the constraints) would allow us to identify
some functional of the phase space coordinate $\tau [h_{ij},
{\pi}^{ij}]$
as a time variable, and would thus allow us to interpret
Equation~(\ref{wdw})
in terms of a quantum evolution with respect to that variable
\cite{KK3}.
However,
we will not allow ourselves the luxury of imagining that we may be so
lucky as to find such an `internal time' but will, instead, use our
embedding variables as a preferred, external, hypertime for the
theory.
As we have seen in Section~\ref{lag}, these variables change the
dynamics of the theory; now we will investigate them at the
Hamiltonian level.

The action $S^R$ can be regarded as simply a standard matter action,
coupling the fields ${\lambda}_A$ and $X^\mu$ to gravity. $F^A$
contains no derivatives of $\bar{g}_{\mu \nu}$, and thus we do not
change
the gravitational velocity-momentum relationships
derived from Equation~(\ref{momh}).
Let us now define an extended phase space ${\cal E} = T^*({\cal C}_h
\times {\cal C}_X)$, where ${\cal C}_X$ is the configuration space
of the embedding variables on $M$. We then have a representation
of the tangent bundle $T({\cal M})$, restricted
to the embedding, through its isomorphic relationship
to the velocity bundle $T({\cal C}_X)$.
Again, we have no momenta conjugate to the
$\lambda_A$ and thus we can simply regard these variables as generating
constraints on the extended phase space, as one might expect from their
r\^{o}le as enforcers of the coordinate conditions.

The Hamiltonian formalism can be expressed in a simpler form
if we define
\begin{equation}{}\label{vdef}
n^\mu = \frac{1}{N} \left( {\dot{X}}^\mu - L_{\bf N} X^\mu \right),
\end{equation}
where $L_{\bf N}$ is the standard Lie derivative in the direction $N^i$
on
$M$. This object has a simple geometric interpretation as being
the normal to the foliation $i$, expressed in the $X^\mu$ coordinates,
in the sense that we have
\begin{equation}{}\label{norm1}
{\bf n} = n^\mu \frac{\partial}{\partial X^\mu} = \frac{1}{N} \left(
\frac{\partial}{\partial t}-N^i \frac{\partial}{\partial x^i} \right),
\ X^\mu_i \frac{\partial}{\partial X^\mu} = \frac{\partial}{\partial
x^i},
\end{equation}
and therefore that Equation~(\ref{admmet}) tells us that
\begin{equation}{}\label{norm2}
{\bf g}_Y \left( {\bf n}, {\bf n} \right) =-1,
\ {\bf g}_Y \left( {\bf n},
		  X^\nu_i \frac{\partial}{\partial X^\nu} \right) = 0
\end{equation}
for all $i$.
This greatly simplifies the form of the Equation~(\ref{gbd}) for
the ADM form of ${\bf g}$ in the $X^\mu$
coordinate system, giving
\begin{equation}{}\label{gbdadm}
\bar{g}^{\mu \nu} = - n^\mu n^\nu + h^{ij} X^{\mu}_i X^{\nu}_j.
\end{equation}
We will now restrict ourselves to $F^A$ functions that correspond
to a physical, unique, embedding, and are thus {\it good} in
the following sense:
\begin{defit}[Good ultralocal coordinate conditions]{}\label{goodc}
A good ultralocal coordinate condition is one with the property
that, for a given specification of $X^\mu$ and ${\bf h}$ on a
hypersurface,
there are at most two real solutions, $\pm {\bf n}_S$, with
$F^A \left( {\bf n}_S \right) =0$.
\end{defit}
Notice that we do not require that $F^A \left( {\bf n}_S \right) =0$
has a solution for all values of $X^\mu$ on the hypersurface; some of
the embeddings will correspond to non-spacelike hypersurfaces in
${\cal M}$ and thus Equation~(\ref{norm2}) will be violated.

Now we are in a position to calculate the momenta conjugate to the
$X^\mu$;
\begin{equation}{}\label{momX}
P_\mu (t,x) = \frac{\delta L^R(t)}{\delta X^\mu_0(t,x)}
	    = \sqrt{h} \lambda_A F^A_{\mu \nu} ({\bf n}) n^\nu,
\end{equation}
where
\begin{equation}{}\label{FMN}
F^A_{\mu \nu} ({\bf n})
= \frac{\partial F^A ({\bf n})}{\partial {\bar g}^{\mu \nu}},
\end{equation}
and where we explicitly indicate the dependence of $F^A_{\mu\nu}$ on
${\bf n}$ whilst not indicating the other
variables for convenience later on.
However, these relations immediately present us with a problem; the
function $$\lambda_A F^A_{\mu\nu}
({\bf n}) n^\nu: T({\cal C}_X) \rightarrow T^*({\cal C}_X)$$
will not always be injective and therefore
not surjective, with the result being that we cannot always invert it
to
retrieve ${\bf n}$, and hence ${\dot{X}}^\mu$, from the momenta.
This means that there must be extra constraints in the theory.
To solve this problem, and to also make it simple to remove the
Lagrange
multipliers from our system, we will solve these extra constraints, and
work with the remaining freedom in $\lambda_A$.
We proceed as follows: Let us define ${\cal C}_\lambda$ to be the
configuration bundle of $\lambda_A$ over $M$, with fiber
${\cal C}_{\lambda x}$. We find that although Equation~(\ref{momX}) is
not generally invertible for ${\bf n}$, as a function of the momenta,
it is generally invertible for $\lambda_A \in {\cal C}_{\lambda x}$,
as a function of the momenta and velocities;
\begin{proovit}[Inversion for Lagrange multipliers]{}\label{invlm}
If the coordinate condition functions $F^A$ are functionally
independent, as functions of ${\bf n}$, then there exists
an open set ${\cal C}'$, of the total configuration bundle,
${\cal C} = {\cal C}_h \times {\cal C}_X \times {\cal C}_\lambda$,
such that the map
\begin{equation}{}\label{Wdef}
W_\mu^A ({\bf n}) = F^A_{\mu\nu} ({\bf n}) n^\mu : {\cal C}_{\lambda x}
\rightarrow T^*_x ({\cal C}_X)
\end{equation}
is invertible for all configurations $c \in {\cal C}'$.
\end{proovit}
{\bf Proof:} Now, we have
\begin{equation}{}\label{Weq}
W^A_{\mu} = -\frac{1}{2} \frac{\partial F^A}{\partial n^\mu},
\end{equation}
and is therefore proportional to the Jacobian matrix for $F^A$ as a
function of ${\bf n}$. Therefore functional independence tells us that
there cannot exist a field $k_A \in {\cal C}_{\lambda}$ such that
$W^A_\mu k_A =0$ for all of ${\cal C}$. However, due to the smoothness
properties of $F^A$, this implies that there must be an open set
${\cal C}'$ that is dense in ${\cal C}$ such that for all $c \in {\cal
C}'$
there does not exist a $k_{Ax} \in T_x ({\cal C}_\lambda)$ with
$W^A_\mu k_{Ax} = 0$. Thus $W^A_\mu$ has no kernel at $c$ and is thus
invertible. {\bf QED}

This theorem therefore allows us to invert Equation~(\ref{momX}) to
get
\begin{equation}{}\label{invmom}
\lambda_A ({\bf n}) = \frac{1}{\sqrt{h}} T^\mu_A ({\bf n}) P_\mu,
\ T^\mu_A ({\bf n}) = \left( W^A_\mu ({\bf n}) \right)^{-1},
\end{equation}
for all $c \in {\cal C}'$ However, for a fixed $P_\mu$, variation
of ${\bf n}$ does not generally produce all values in ${\cal
C}_\lambda$,
but only those in a subspace, denoted by ${\cal C}^{\|}_\lambda$. Thus,
the values of $\lambda_A$ are constrained to lie in this space, and
thus
we can label these physical values by a new Lagrange multiplier,
${\bf \bar{n}}$, with the substitution
\begin{equation}{}\label{newlag}
\lambda_A ({\bf \bar{n}}) = \lambda_A ( {\bf n} \rightarrow
{\bf \bar{n}} ) = \frac{1}{\sqrt{h}} T^\mu_A ({\bf \bar{n}}) P_\mu.
\end{equation}
The $n$ components of ${\bf \bar{n}}$ are, of course, an overcounting
of the values in ${\cal C}^{\|}_\lambda$; let
${\cal C}_{\bar{n}}$ be the configuration space for ${\bf \bar{n}}$,
and
let us define the following equivalence relation:
\begin{defit}[Equivalence on ${\cal C}_{\bar{n}}$]{}\label{equivn}
We say that ${\bf \bar{n}},{\bf \bar{n}'} \in {\cal C}_{\bar{n}}$
are equivalent, denoted by ${\bf \bar{n}} \sim {\bf \bar{n}'}$,
if and only if $\lambda_A ({\bf \bar{n}}) = \lambda_A ({\bf
\bar{n}'})$.
We denote the equivalence classes of ${\cal C}_{\bar{n}} / \! \sim$ by
$[{\bf \bar{n}}] \subseteq {\cal C}_{\bar{n}}$.
\end{defit}
We can then obviously see that ${\cal C}^{\|}_\lambda$ is isomorphic to
the quotient space ${\cal C}_{\bar{n}} / \! \sim$.
However, for technical reasons, we will have to define a finer
equivalence
relationship on ${\cal C}_{\bar{n}}$, to allow full inversion
of the velocity-momentum relations; in general, $[{\bf \bar{n}}]$ will
not be a single path component of ${\cal C}_{\bar{n}}$ and thus we will
split up these equivalence classes into a discrete
union of path components and define a new relation as follows:
\begin{defit}[Local equivalence on ${\cal
C}_{\bar{n}}$]{}\label{lequivn}
We say that ${\bf \bar{n}},{\bf \bar{n}'} \in {\cal C}_{\bar{n}}$
are locally equivalent, denoted by ${\bf \bar{n}} \leftrightarrow
{\bf \bar{n}'}$,
if and only if there exists a continuous path ${\bf \bar{m}}(s) \in
{\cal C}_{\bar{n}}$ with $s \in [0,1]$, ${\bf \bar{m}}(0) = {\bf
\bar{n}}$,
${\bf \bar{m}} (1) = {\bf \bar{n}'}$, and ${\bf \bar{m}}(s) \sim
{\bf \bar{n}}$ for all $s$.
We denote the equivalence classes of
${\cal C}_{\bar{n}} / \! \leftrightarrow$ by
$[{\bf \bar{n}}]_L \subseteq {\cal C}_{\bar{n}}$.
\end{defit}
Obviously, ${\cal C}_{\bar{n}} / \! \leftrightarrow$ is a covering
space
of ${\cal C}_{\bar{n}} / \! \sim$ and therefore of ${\cal
C}^{\|}_\lambda$.
We can now invert
the velocity-momentum relations on this
covering space to arrive at the Hamiltonian
form of $S^R$ restricted to ${\cal C}_{\bar{n}} / \! \leftrightarrow$;
\begin{equation}{}\label{srham}
S^R \left[ h_{ij}, P_\mu, X^\nu, N, N^i, [{\bf \bar{n}}]_L \right] =
\int_{\cal M} dt \, d^3x \left( \dot{X}^\mu P_\mu - NH^R - N^i H^R_i
\right),
\end{equation}
where
\begin{equation}{}\label{hr}
H^R = \bar{n}^\mu P_\mu + \frac{1}{2} T^\mu_A ({\bf \bar{n}})
F^A ({\bf \bar{n}}) P_\mu,
\end{equation}
\begin{equation}{}\label{hri}
H^R_i = X^\mu_i P_\mu.
\end{equation}
to verify the consistency of this equation, we must check that it
is invariant to the choice of representative of ${\bf \bar{n}'}
\in [{\bf \bar{n}}]_L$;
\begin{proovit}[Consistency of Hamiltonian action on
covering space]{}\label{ConHam}
If ${\bf \bar{n}},{\bf \bar{n}'} \in {\cal C}_{\bar{n}}$ and
${\bf \bar{n}} \leftrightarrow {\bf \bar{n}'}$ then we have $H^R(
{\bf \bar{n}}) = H^R( {\bf \bar{n}'})$.
\end{proovit}
{\bf Proof:} We obviously only need to verify that the action is
invariant under all small perturbations within $[{\bf \bar{n}}]_L$. It
is
a simple matter to show that
\begin{equation}{}\label{HRder}
\frac{\partial H^R}{\partial \bar{n}^\mu} = \frac{1}{2} \sqrt{h}
\frac{\partial \lambda_A}{\partial \bar{n}^\mu} F^A.
\end{equation}
Now let ${\bf \bar{n}} \rightarrow {\bf \bar{n}} + \delta {\bf k}$,
with $\delta {\bf k} \in {\cal C}_{\bar{n}}$ small and
${\bf \bar{n}} \sim {\bf \bar{n}} + \delta {\bf k}$. Thus we
have
\begin{equation}{}\label{nulv}
\frac{\partial \lambda_A}{\partial \bar{n}^\mu} \delta k^\mu = 0,
\end{equation}
and therefore $H^R({\bf \bar{n}} + \delta {\bf k}) =  H^R({\bf
\bar{n}})$.
Hence the result follows, and $S^R$ is a well-defined functional on
${\cal C}_{\bar{n}} / \! \leftrightarrow$. {\bf QED}

We are now in a position to remove $[{\bf \bar{n}}]_L$ from the
action by solving the dynamics of the covering space extended theory
for it;
\begin{proovit}[Removal of Lagrange multipliers]{}\label{rlm}
Variation of $[{\bf \bar{n}}]_L \in {\cal C}_{\bar{n}} / \!
\leftrightarrow$
in the action $S^R[h_{ij},P_\mu, X^\mu, N, N^i, [{\bf \bar{n}}]_L]$ on
the covering space extended theory gives us
$H^R = \bar{n}^\mu_S P_\mu$, where ${\bf \bar{n}}_S \in {\cal
C}_{\bar{n}}$
is one of the $P_\mu$ independent solutions of $F^A ({\bf \bar{n}}_S) =
0$.
This solution is valid for all of ${\cal C}$ if we require continuity
of $H^R$.
\end{proovit}
{\bf Proof:} From Equation~(\ref{HRder}) in Theorem~\ref{ConHam}
we have
\begin{equation}{}\label{miniS}
\frac{\delta S^R}{\delta \bar{n}^\mu} = 0 \Leftrightarrow
\frac{\partial H^R}{\partial \bar{n}^\mu} = 0 \Leftrightarrow
\frac{\partial \lambda_A}{\partial \bar{n}^\mu} F^A = 0.
\end{equation}

This defines a discrete set of surfaces
in ${\cal C}_{\bar{n}}$ each corresponding
to a single point $[{\bf \bar{n}}_e]_L \in
{\cal C}_{\bar{n}} / \! \leftrightarrow$,
from Theorem~\ref{ConHam}. However, these solutions include
those of the form $[{\bf \bar{n}}_S]_L$, with
$F^A({\bf \bar{n}}_S) =0$. This solution is valid for all of
${\cal C}'$, but the closure of ${\cal C}'$ is ${\cal C}$, and
thus, by continuity, the result follows. {\bf QED}

Now,  $\bar{n}^\mu_S$ is specified precisely, up to sign,
by $F^A=0$ as long as the coordinate conditions are
`good' in the sense defined earlier; we will therefore pick out
a single component, $[{\bf \bar{n}}_S]_L \subseteq [{\bf \bar{n}}_S]$,
and thus our variation gives a well-defined couplet of solutions
on ${\cal C}^{\|}_\lambda$.
Due to the relationship between $\bar{n}^\mu$ and $n^\mu$, and hence
$X^\mu_0$, $\bar{n}^\mu_S$ will be independent, as a vector, from the
$n-1$ vectors $X^\mu_i$. Thus, we are motivated to define
\begin{equation}{}\label{Ydef}
Y^\mu_{\nu} = \left\{ \begin{array}{ll}
		   \bar{n}^\mu_S & \mbox{if $\nu = 0$} \\
		   X^\mu_\nu & \mbox{if $\nu > 0$} \end{array} \right.
\end{equation}
which is an object defined purely on the bundle $T^*({\cal C}_h \times
{\cal C}_X)$ and is independent of $P_\mu$. Let us now define
$N^\mu = (N,N^i),\ H^G_\mu=(H^G, H^G_i)$ and $H^R_\mu =(H^R,H^R_i)$.
In this notation
we have the following simple form for the embedding action without
Lagrange multipliers:
\begin{equation}{}\label{Srnice}
S^R[h_{ij}, X^\mu, P_\nu, N^\sigma] = \int_{\cal M} dt \, d^3 x \left(
\dot{X}^\mu P_\mu - N^\mu H^R_\mu \right),
\end{equation}
where
\begin{equation}{}\label{HRnewd}
H^R_\mu = Y^\nu_\mu P_\nu.
\end{equation}
The total action for the theory is thus
\begin{equation}{}\label{STnice}
S[h_{ij}, \pi^{lm}, X^\mu, P_\nu, N^\sigma] = \int_{\cal M} dt \, d^3 x
\left(
\dot{X}^\mu P_\mu + \dot{h}_{ij} \pi^{ij} - N^\mu H_\mu \right),
\end{equation}
and $H_\mu$ is given by
\begin{equation}{}\label{Hdef}
H_\mu = H_\mu^G + H_\mu^R = H_\mu^G + Y^\nu_\mu P_\nu,
\end{equation}
with the $N^\mu$ variation giving us constraints $H_\mu=0$
on the embedding-extended phase space.
To bring out the final form of the action, let us now define
$Q^\mu_\nu$ to be the inverse of $Y^\mu_\nu$, $Q^\mu_\nu Y^\nu_\sigma
= \delta^\mu_\sigma$. We can then define a new, equivalent, set of
constraint functions by
\begin{equation}{}\label{newcon}
\Pi_\mu = Q^\nu_\mu H_\mu = P_\mu + Q^\nu_\mu H^G_\nu,
\end{equation}
and redefine the lapse and shift to get
\begin{equation}{}\label{STnicer}
S[h_{ij}, \pi^{lm}, X^\mu, P_\nu, N^\sigma] = \int_{\cal M} dt \, d^3 x
\left(
\dot{X}^\mu P_\mu + \dot{h}_{ij} \pi^{ij} - \bar{N}^\mu \Pi_\mu
\right),
\end{equation}
which corresponds to $N^\mu = \bar{N}^\nu Q^\mu_\nu$. The variation of
these new lapse and shift functions then gives us $\Pi_\mu=0$.
It is this Hamiltonian formulation that allows us to construct a
representation of spacetime diffeomorphisms, and which contains a
copy of the standard Einstein geometrodynamics.

We have now
made contact with the formalism developed by Isham, Kucha\v{r}
and Torre \cite{DIFIK2,KT1} and
will now proceed to list the generic behavior present, which
they derived for the specific case of the Gaussian coordinate
condition.
The proofs are identical, and a reader familiar with the embedding
formalism
may now move on to the discussion section. We present them here purely
for completeness.
This formalism, like standard geometrodynamics, is constrained so
that the physical dynamics lie on $\Gamma_e \subset T^*({\cal C}_h
\times {\cal C}_X)$, where $\Pi_\mu(\Gamma_e)=0$. If we now define
$h_\mu = Q^\nu_\mu H^G_\nu$, the `unprojected' pure gravity constraint
functions, then $\Gamma_e$ also contains a copy of $\Gamma$, the
standard constraint surface, determined by
$h_\mu(\Gamma)=0=H^G_\mu(\Gamma)$.
The latter surface corresponds to the intersection of the
surface $\Gamma_H \subset T^*({\cal C}_h \times {\cal C}_X)$, defined
by
$P_\mu(\Gamma_H)$ = 0, and $\Gamma_e$.
The total Hamiltonian on $T^*({\cal C}_h \times {\cal C}_X) \times
{\cal C}_{\bar{N}}$, where ${\cal C}_{\bar{N}}$ is simply the
configuration
space for $\bar{N}^\mu$, is
\begin{equation}{}\label{toth}
{\bf H}_T = \int_M d^3x \bar{N}^\mu \Pi_\mu.
\end{equation}
For the dynamics to be consistent, all that we would require would
be that $\dot{\Pi}_\mu = \{ \Pi_\mu, {\bf H}_T \}$ vanishes weakly
(vanishes
on $\Gamma_e$), and thus that the constraints are propagated by
the dynamics induced by ${\bf H}_T$. However, we have a far stronger
result, namely
\begin{proovit}[Abelian constraint algebra]{}\label{aca}
$\{ \Pi_\mu, \Pi_\nu \}$ vanishes strongly.
\end{proovit}
{\bf Proof:} Isham and Kucha\v{r} originally proved this result with a
long
and complicated calculation \cite{DIFIK2}, but it can be derived from a
simple
argument \cite{KT1}; The original constraint functions, $H_\mu$, obey
the Dirac algebra, given by Equations~(\ref{ccr1}) to~(\ref{ccr3}),
and thus $\{H_\mu, H_\nu\}$ vanishes weakly. Furthermore, the new
constraints
are equivalent to the old ones and thus $\{\Pi_\mu,\Pi_\nu\}$
weakly vanishes. However $\{\Pi_\mu,\Pi_\nu\}$ does not include
any terms containing $P_\mu$, due in particular to
the independence of $Q^\mu_\nu$ on the momenta,
and hence the value of the commutator
cannot depend of the value of the constraint functions, and thus
we must have $\{\Pi_\mu,\Pi_\nu\}=0$ strongly. {\bf QED}

We can thus proceed to calculate the dynamics of this general IKU
embedding-extended theory by evaluating the rest of the brackets. We
get the following evolution equations:
\begin{equation}{}\label{Xdot}
\dot{X}^\mu(x) = \{ X^\mu(x), {\bf H}_T \} = \bar{N}^\mu (x),
\end{equation}
\begin{equation}{}\label{hdot}
\dot{h}_{ij}(x) = \{ h_{ij}(x), {\bf H}_T \} = \int_M d^3x'
				     N^\mu(x') \{ h_{ij}(x),
				     H^G_\mu(x') \},
\end{equation}
\begin{equation}{}\label{Pdot}
\dot{P}_\mu (x) = \{ P_\mu (x), {\bf H}_T \} = \int_M d^3x' \bar{N}^\nu
(x')
				       \{ P_\mu(x), Q^\sigma_\nu (x')
				       \}
				      H^G_\sigma (x'),
\end{equation}
and
\begin{eqnarray}{}\label{pidot}
\dot{\pi}^{ij} (x) & = &\{ \pi^{ij}, {\bf H}_T \}   \nonumber \\
 & = & \int_M d^3x' N^\mu (x') \{ \pi^{ij}(x), H^G_\mu(x') \} +
 \nonumber \\
 & &               \int_M d^3x' \bar{N}^\nu (x') \{ \pi^{ij}(x),
		     Q^\mu_\nu(x') \} H^G_\mu (x').
\end{eqnarray}
Where we see the relation ${\bf n} = {\bf \bar{n}}_S$ showing up
in Equation~(\ref{Xdot}), thus verifying the consistency of
the dynamics.
The coupling of the embedding field to gravity shows up in the second
term in $\dot{\pi}^{ij}$, which vanishes when $H^G_\mu=0$. We have
the following generalization of the work by
Isham and Kucha\v{r} \cite{DIFIK2}:
\begin{proovit}[Einstein-Hilbert sector]{}\label{EHS}
The space $\Gamma_e \cap \Gamma_H$ is preserved under the dynamical
evolution generated by ${\bf H}_T$. The resulting dynamics are those of
pure Einstein gravity.
\end{proovit}
{\bf Proof:} Let the system start with $\Pi_\mu = H^G_\mu = P_\mu =
0$.
Equation~(\ref{Pdot}) gives us $\dot{P}_\mu = 0$, and Theorem~\ref{aca}
gives us $\dot{\Pi}_\mu =0$. Hence $\dot{h}_{\mu} =0$, and thus
$h_\mu = 0 = H^G_\mu$ for all of the resulting trajectory in phase
space.
Equations~(\ref{hdot}) and~(\ref{pidot}) then tell us that $h_{ij}$
and $\pi^{ij}$ propagate according to the standard pure
geometrodynamics
evolution equations. {\bf QED}

Thus, once we have $P_\mu =0$, no internal gravitational dynamics
can couple to the embedding variables, and pure gravity results.
Therefore, all that we have left in our advertized itinerary is
the following:
\begin{proovit}[Representation of spacetime
diffeomorphisms]{}\label{std}
The space
$T^*({\cal C}_h \times {\cal C}_X)$ contains a full homomorphic
representation of the action of the Lie algebra of the spacetime
diffeomorphism group.
\end{proovit}
{\bf Proof:} The Lie algebra of ${\rm Diff}({\cal M})$ is just the
space of complete vector fields on ${\cal M}$. Therefore, let
$${\bf u} \left( X^\alpha (x) \right) = u^\mu
\frac{\partial}{\partial X^\mu} \in {\rm LDiff} ({\cal M}), $$ then we
define
\begin{equation}{}\label{stdrep}
{\bf \Pi} ({\bf u}) = \int_M d^3x u^\mu {\Pi}_\mu.
\end{equation}
Therefore, we have
\begin{equation}{}\label{com1}
\{ {\bf \Pi} ({\bf u}), {\bf \Pi} ({\bf v}) \} =
\int_M d^3x \Pi_\mu \left( v^\nu \frac{\partial u^\mu}{\partial X^\nu}
	 -  u^\nu \frac{\partial v^\mu}{\partial X^\nu} \right),
\end{equation}
Where we have used
\begin{equation}{}\label{com2}
\left\{ u^\mu \left( X^\alpha(x) \right), \Pi_\nu (x') \right\}
= \delta (x-x') \frac{\partial u^\mu (x)}{\partial X^\nu (x')}.
\end{equation}
However, the Lie bracket $[{\bf u}{\bf v}]$ on ${\rm LDiff}({\cal M})$
is
simply
\begin{equation}{}\label{com3}
[{\bf u}{\bf v}] = - [{\bf u},{\bf v}] =
\left( v^\nu \frac{\partial u^\mu}{\partial X^\nu}
	 -  u^\nu \frac{\partial v^\mu}{\partial X^\nu} \right)
\frac{\partial}{\partial X^\mu},
\end{equation}
and therefore
\begin{equation}{}\label{com4}
\{ {\bf \Pi} ({\bf u}), {\bf \Pi} ({\bf v}) \} =
{\bf \Pi} \left( [{\bf u}{\bf v}] \right),
\end{equation}
and thus we have our desired result. {\bf QED}

We therefore have our desired Hamiltonian theory containing
a representation of the Lie algebra of spacetime diffeomorphisms
as well as a sector that obeys standard pure geometrodynamics.

\section{Discussion}{}\label{disc}
To end this paper, we shall mention a few important issues that
crop up in the
IKU formalism. The primary motivation behind this construction is the
hope of formulating a consistent theory of quantum geometrodynamics.
The quantum mechanical formalism is achieved by replacing the phase
space coordinates with operators with the standard canonical rules,
in which configuration variables are replaced by
\begin{equation}{}\label{canquant1}
X^\mu \rightarrow \hat{X}^\mu = X^\mu \times,\ h_{ij}
\rightarrow \hat{h}_{ij} = h_{ij} \times,
\end{equation}
and the momenta are replaced by
\begin{equation}{}\label{canquant2}
P_\mu \rightarrow \hat{P}_\mu = -i
\frac{\delta}{\delta X^\mu},\ \pi^{ij} \rightarrow \hat{\pi}^{ij} =
-i \frac{\delta}{\delta h_{ij}},
\end{equation}
acting on an embedded wavefunction $\psi({\bf h}, X^\mu)$.
Our first major improvement is that the IKU theory describes a
wavefunction propagating from one spacelike surface to another,
labeled by the preferred hypertime,
\begin{equation}{}\label{ikueq}
i\frac{\delta}{\delta X^\mu} \psi({\bf h},X^\mu) =
	  \hat{h}_\mu \psi({\bf h},X^\mu),
\end{equation}
in comparison to the standard QGD Wheeler-DeWitt equation, given
by Equation~(\ref{wdw}), which contains no mention of time. Thus the
IKU construction gives us a natural multifingered Schr\"odinger
equation. The second point is that the quantum theory then carries
a representation of ${\rm LDiff} ({\cal M})$, as long as
Theorems~\ref{aca} and~\ref{std} have quantum equivalents. The latter
condition
is equivalent to asking for an ordering of the operators in
$\hat{\Pi}_\mu$ so that
\begin{equation}{}\label{qcomr}
[ \hat{\Pi}_\mu, \hat{\Pi}_\nu ] =0,
\end{equation}
where $[,]$ denotes the standard group commutator. Unfortunately,
achieving
this can be a significant technical problem, as discussed by
Kucha\v{r} and Torre \cite{KT1},
and may limit the applicability of this work in QGD. The most obvious
remaining problem is that the IKU states corresponding
to Equation~(\ref{ikueq}) do not correspond
classically to pure geometrodynamics, but to the
theory given by Equation~(\ref{newE}). If we wish to recover
Einstein gravity, we must further require $\hat{P}_\mu \psi=0$,
which corresponds to $\psi$ being independent of hypertime,
and seemingly returns us to the problematic world of the
Wheeler-DeWitt equation. Thus the `real' time variable becomes
inaccessible, and we, trapped inside the universe, cannot use it
in the so-called Heraclitian sense (cf. Ref. \cite{BILLBOB1}) to
provide an ordering of events for us.
There are two possible exits to this problem;
The first is that the embedding fields may correspond to actual
physical fields, as discussed by Kucha\v{r} and Torre \cite{KT1,KT2},
and thus
the constraint $\hat{P}_\mu \psi=0$ is not needed. The second exit
is that we can, in principle, construct $\hat{P}_\mu \psi=0$
states out of a superposition of general solutions to the IKU
equation, and may thus be able to use the external hypertime
as an aid to interpretation. This latter point has been partially
discussed by Halliwell and Hartle \cite{HHOC,HGS}.

The other issue that the Author wishes to raise is the generic
nature of this construction; Although we have confined ourselves to
Einstein gravity, it is obvious that the construction is valid for any
theory
of quantum gravity that has constraints similar to the Dirac algebra.
We can quite easily couple matter fields, or even make canonical
transformations on the geometrodynamic phase space, without changing
our results.
In fact, the formalism is only dependent on our ability to construct
the `unprojected' constraints given by Equation~(\ref{newcon})
in such a way that $P_\mu$ appears through the linear term only,
and does not appear in $h_\mu$. It is
this essential property that leads to the representation of spacetime
diffeomorphisms, as well as to the existence of a consistent,
classical,
copy of the original theory inside the extended phase space.
\section{Conclusion}{}\label{conc}
We have constructed a general version of the IKU
embedding-extended formalism for quantum geometrodynamics that
solves the problem of time whilst exchanging it for a collection of
extra
physical fields. The classical theory includes a copy of pure
geometrodynamics, but one cannot be sure whether
it is possible for this sector to
be realized consistently in the quantum theory. The classical
theory contains a representation of the Lie algebra of the spacetime
diffeomorphism group, and this result may also extend to the
quantum theory if certain operator ordering problems can be solved.
\section*{Acknowledgments}{}\label{acks}
I would like to warmly thank Diane Delap for discussions and support.
This work was supported
by the Natural Sciences and Engineering Research Council of Canada.

\end{document}